\documentclass{nime-alternate} 

\usepackage{anonymize} 		   
\usepackage{balance}
\usepackage[utf8]{inputenc}
\usepackage{booktabs}
\usepackage[hidelinks]{hyperref}

\usepackage{capt-of}

\begin{document}


\conferenceinfo{NIME'23,}{31 May–2 June, 2023, Mexico City, Mexico.}

\title{LoopBoxes -- Evaluation of a Collaborative Accessible Digital Musical Instrument}
%
%
%
\label{key}
%

\numberofauthors{5} 
\author{
\alignauthor 
\anonymize{Andreas F{\"o}rster}\\
            \affaddr{\anonymize{Technische Universit{\"a}t Berlin}}\\
           \affaddr{\anonymize{University of Furtwangen}}\\
           \email{\anonymize{andreas@imui.org}}
\alignauthor 
\anonymize{Alarith Uhde}\\
         \affaddr{\anonymize{University of Siegen}}\\
        \affaddr{\anonymize{Siegen, Germany}}\\
        \email{\anonymize{alarith.uhde@uni-siegen.de}}
\alignauthor
\anonymize{Mathias Komesker}\\
            \affaddr{\anonymize{imui e.V.}}\\
           \affaddr{\anonymize{Cologne, Germany}}\\
           \email{\anonymize{mathias@imui.org}}
\and
\alignauthor
\anonymize{Christina Komesker}\\
           \affaddr{\anonymize{imui e.V.}}\\
           \affaddr{\anonymize{Cologne, Germany}}\\
           \email{\anonymize{christina@imui.org}}
\alignauthor
\anonymize{Irina Schmidt}\\
        \affaddr{\anonymize{imui e.V.}}\\
           \affaddr{\anonymize{Cologne, Germany}}\\
        \email{\anonymize{irina\_schmidt88@web.de}}
}

\maketitle


\begin{figure*}[htbp]
\centering
  \includegraphics[width=\linewidth]{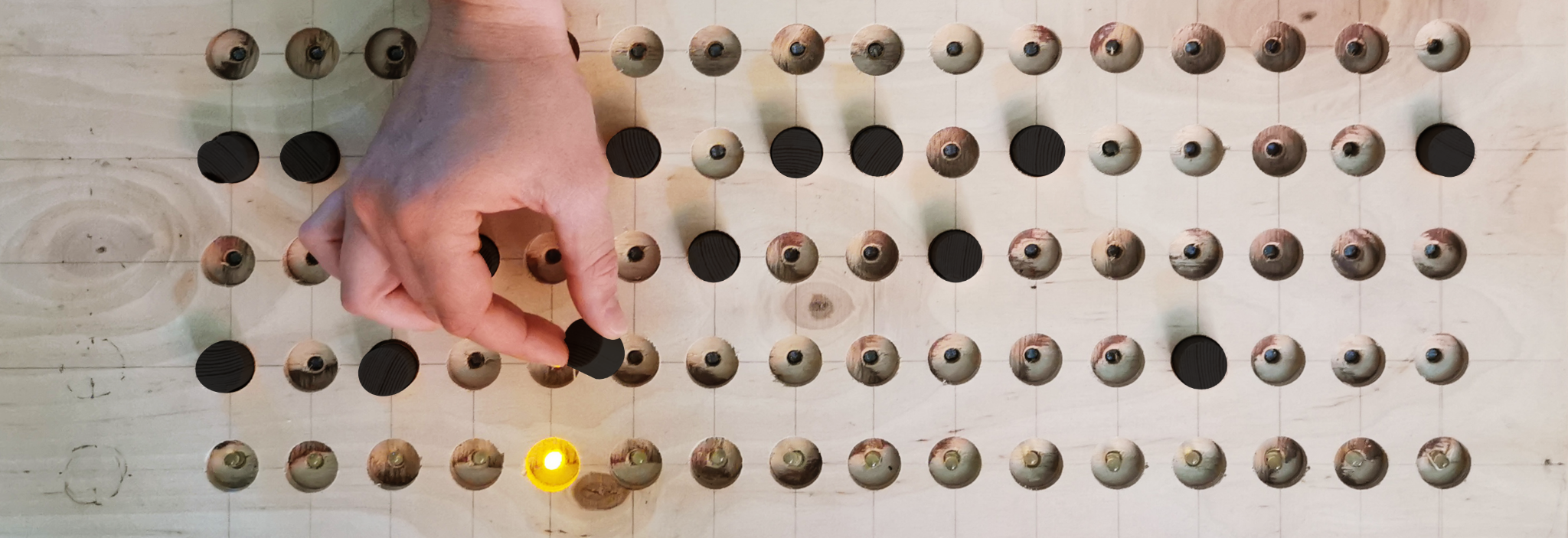}
  \caption{First prototype of the step sequencer module}
  \label{fig:teaser}
\end{figure*}

\begin{abstract}
  LoopBoxes is an accessible digital musical instrument designed to create an intuitive access to loop based music making for children with special educational needs (SEN). This paper describes the evaluation of the instrument in the form of a pilot study during a music festival in Berlin, Germany, as well as a case study with children and music teachers in a SEN school setting. We created a modular system composed of three modules that afford single user as well as collaborative music making. The pilot study was evaluated using informal observation and questionnaires ($n = 39$), and indicated that the instrument affords music making for people with and without prior musical knowledge across all age groups and fosters collaborative musical processes. The case study was based on observation and a qualitative interview. It confirmed that the instrument meets the needs of the school settings and indicated how future versions could expand access to all students.
  especially those experiencing complex disabilities. In addition, out-of-the-box functionality seems to be crucial for the long-term implementation of the instrument in a school setting.
 
\end{abstract}





\section{Introduction}

Over the past years, there has been a growing trend towards digital musical instruments (DMIs) specifically designed for people with disability experience \cite{Frid2019Accessible}. 
Despite this growing interest, DMIs specifically designed to be used by children with special educational needs (SEN) in school settings are still rare. One problem is that most available instruments are designed for single-user interaction \cite{Frid2019Accessible}. However, SEN school settings typically afford playing music in groups, where teachers have to orchestrate the different instruments, and manage and maintain them in real time. Accordingly, in a recent large-scale survey conducted in Germany, teachers expressed a preference for simple, modular, and interactive units that can be played together, rather than complex individual instruments \cite{frontiers23}. 

Thus, we set out to design an accessible set of DMIs that fit with the needs of SEN school settings by focusing on modularity, collaborative music making, tangible interaction, and freely available open source components. This paper describes the design approach as well as two studies that informed the iterative design of the DMIs. 

Our goal was not only to make the instrument suitable for use in SEN settings, but also to develop an instrument that is aesthetically appealing for both individuals with and without disabilities, to counter the trend of considering DMIs for this area as a ``substandard practice'' \cite[p. 3]{beatpaper}.
Thus, we conducted a pilot study at a music festival with a mixed audience. In the second step, we evaluated the instrument at a SEN school.

\section{Related Work}

There are numerous DMIs that have been developed for people with disability experiences. These instruments are also referred to as accessible DMIs (ADMIs) \cite{Frid2019Accessible}. Commercially available options include touchless instruments such as \textit{Soundbeam},\footnote{
\url{https://www.soundbeam.co.uk/}
} haptic controllers like \textit{Skoog}\footnote{
\url{https://skoogmusic.com/}
}, breath controlled instruments like the \textit{Magic Flute}\footnote{
\url{https://mybreathmymusic.com/en/magic-flute}
}, or instrument that use mixed modalities like \textit{Touch Chord}\footnote{
\url{https://www.humaninstruments.co.uk/}
}, which is both breath- and touch-controlled. Additionally, there are developments that are not (yet) commercially available, including bespoke instruments like \textit{Kellycaster}\footnote{
\url{https://www.drakemusic.org/technology/instruments-projects/the-kellycaster/}
}, an adapted guitar that allows chord selection by using a keyboard or AI based instruments \cite{boris}. There are also tangible MIDI controllers designed for education \cite{csme20} and toolkits that allow music teachers to create their own customized music interfaces \cite{ParkeWolfe2019}. The problem for widespread use is that literature shows that for a large part of music teachers such possibilities are not practical, since instruments not working plug-and-play are hardly used \cite{davis}.




Similar developments to ours have been presented in the literature. Compared to our development, which is intended to combine both controller and sound generation in the long run, most of them do not work in standalone mode \cite{bubblegum, bearing}.
Moreover, others are too large for daily use in schools (GRIDI\footnote{\url{https://www.gridi.info/}}), take more experimental approaches to music making \cite{drumtop} or use rather complex visual tracking technologies that seem less suitable for everyday use \cite{marbletrack}.

A similar approach to collaborative music making in a school setting has been presented in \cite{NIME21_63}. Here, the instrument design was mainly related to well-known acoustic instruments (e.g., \textit{the piano} or \textit{the drum}). However, direct references to acoustic musical instruments can be problematic. On the one hand, they are typically oriented towards existing concepts taken from the original instrument, but reduce their complexity. In SEN settings, this is sometimes associated with a deficit-oriented view (``inferior instruments'' or ``no `real' instruments'') \cite{gerlandEchte2019, beatpaper}, or as undesired deviations from cultural notions of ``normality'' (e.g., `normally' a piano is not played with a button) \cite{forsterhci}. In contrast, deviating from traditional instrument metaphors allows designers to instead focus on the unique possibilities of DMIs, while being able to tailor the interaction to the abilities of the children. To address this issue, we developed instruments that use interaction metaphors common in electronic music making. Further, we decided to test our instrument with a mixed audience as well as with a group of children in an SEN setting to ensure the instruments are aesthetically appealing to a broad user group and thus afford inclusive music making. 

Previous work emphasizes that people with disabilities need to be involved in the development of technology for people with disabilities in order to understand their individual needs \cite{bell_hacking}. In addition, in a school setting, other stakeholders are significant: the teachers as well as the broader school as an institution with specific pedagogical requirements. Those requirements were investigated in a comprehensive survey of SEN schools \cite{forsterhci, frontiers23}, which revealed that a major barrier to using DMIs is a general lack of knowledge about their possibilities. In addition, there are basically two different design approaches. Either the development of generic instruments that are as flexible as possible or the development of many simple instruments with clear affordances that address different access needs. The survey showed that for school use the second approach is preferred by the teachers. Therefore, we decided to develop different modules in a first step, which enable making music together and exemplify possibilities of DMIs. These served as a starting point in the course of our project. In the further course, we want to continue expanding the modules and adapt them to the individual needs of the students. The goal is to provide a larger selection of instruments from which teachers can choose the most suitable ones. Furthermore, the survey revealed a list of requirements from the perspective of music teachers in SEN education. These include: 

\begin{itemize}
    \item manageability (ease-of-use, intuitive interaction)
    \item good sound quality
    \item financial viability
    \item robustness
    \item facilitating discovery learning and aesthetic experiences
    \item multimodality (multisensory and direct feedback)
    \item to foster collaborative music making
    \item appealing design aesthetics
\end{itemize}

The initial design of LoopBoxes was informed by these requirements, as described in the next section.


\section{Initial Design}



\begin{figure*}[htbp]
  \centering
  \includegraphics[width=0.8\textwidth]{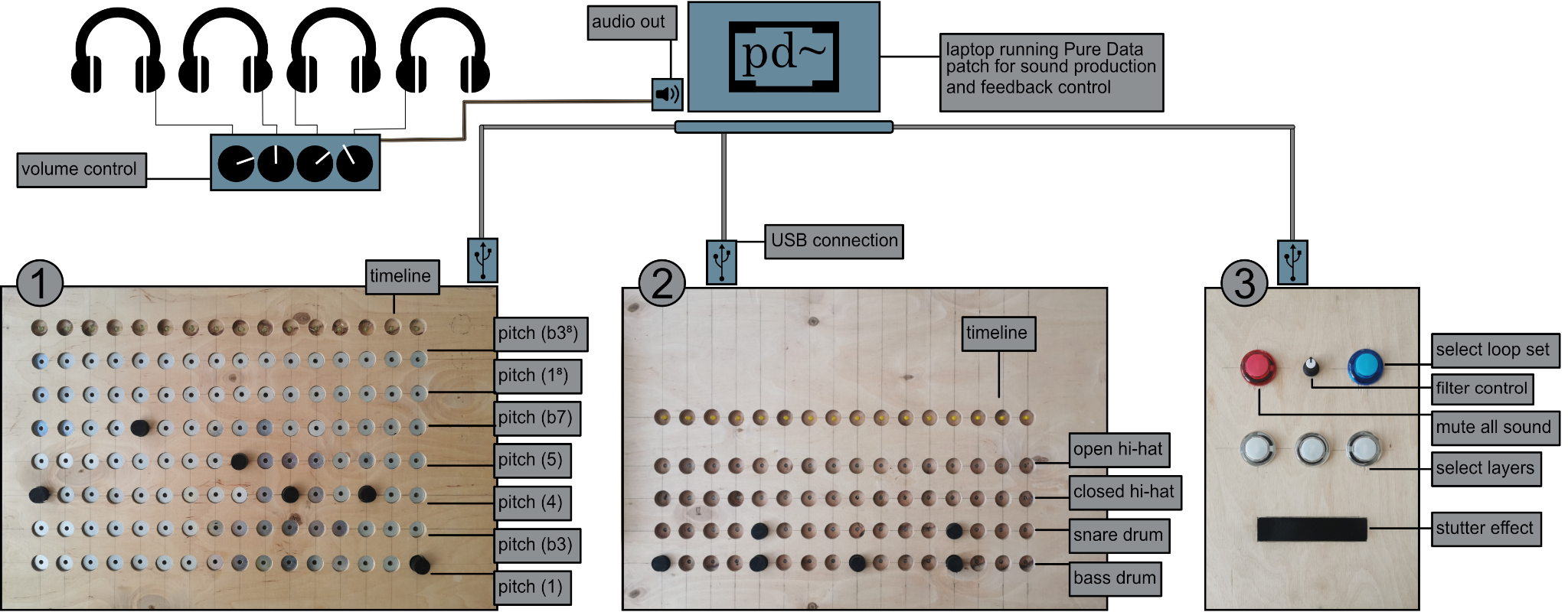}
  \caption{Schematic View of the three modular instruments.}
  \label{fig:instrument}
\end{figure*}

Figure \ref{fig:instrument} shows a schematic view of the instrument. To enable collaborative music making, the instrument consists of three modules with distinct functions. Two step sequencer modules that also work as stand-alone instruments (see \cite{NIME21_4} for a detailed description of the design) and afford
the creation of (1) a melodic line and (2) a drum beat. Module three controls the layers of a background loop, a low- and high-cut filter, a stutter effect, and it mediates the overall interaction.

All modules communicate via serial port with a laptop running a Pure Data patch. The sound was distributed using a headphone mixer enabling each musician to control the volume with a separate control knob during the pilot study. In the SEN school the sound was output using the sound system that was available in the music room.



\subsection{Loop-based Music}

We chose a loop-based design, because this way of music making is the foundation of many popular musical genres (e.g., Electronic Dance Music (EDM)), and we assumed it would connect to the children's listening experiences. In addition, the non-linear, repetitive approach to music making might intrinsically increase accessibility \cite{beatpaper}.
The instrument provides musical loops from different popular music genres that we hope correspond to the users’ listening experiences and contribute to the fostering of aesthetic experiences which, following Christopher Wallbaum \cite{wallbaum}, should be the purpose of music education. 


\subsection{Interface Design}

Current music education research often focuses on the use of touchscreen-based interfaces like the iPad \cite{godauInklusion2018}. Besides many advantages (e.g., off-the-shelf availability, standardization), such devices can also be perceived as a barrier by people with cognitive disabilities \cite{donnerKulturpadagogische}. In addition, the current, easily available touch-based systems typically rely on (at least partially) closed-source solutions. This increases the risk of vendor lock-in and high costs.

As an alternative, we decided to develop the three modules using tangible interfaces, which have a particularly high potential to facilitate accessibility \cite{NIME21_4}. They support intuitive interaction by setting restrictions to distinct functions. In addition, the use of buttons to create direct feedback (cause and effect) is very common in SEN settings and can facilitate access for people with cognitive disabilities \cite{maggeezigoarticla}. 
Furthermore, the modular design was chosen to foster collaborative music making and aesthetic communication \cite{wallbaum} among the users.

The two step sequencers (left and middle in figure \ref{fig:instrument}) provide primary haptic feedback\footnote{
Primary haptic feedback encompasses the haptic qualities of the instrument/material itself as opposed to secondary feedback describing feedback that might be created e.g., by a vibration motor.} by using wooden blocks that are put in a preconfigured arrangement of holes (see Figure \ref{fig:teaser} \& \ref{fig:instrument}). Visual feedback is provided by 16 LEDs indicating the current step that is being played.

 Module three consists of five coloured buttons that provide visual feedback by built-in LEDs. Three white buttons control different layers of the background loop (bassline, chords, atmosphere). The blue button skips to the next set of background loops (altogether, four sets are provided from different musical genres) and the red button mutes the entire system. The touch slider controls a `stutter-effect' that can be accelerated or decelerated by moving the finger to the right or left. The tempo of the stutter effect is represented visually by the speed of blinking LEDs (also at the two step sequencer modules).

\begin{figure}[htbp]
  \centering
  \includegraphics[width=.8\linewidth]{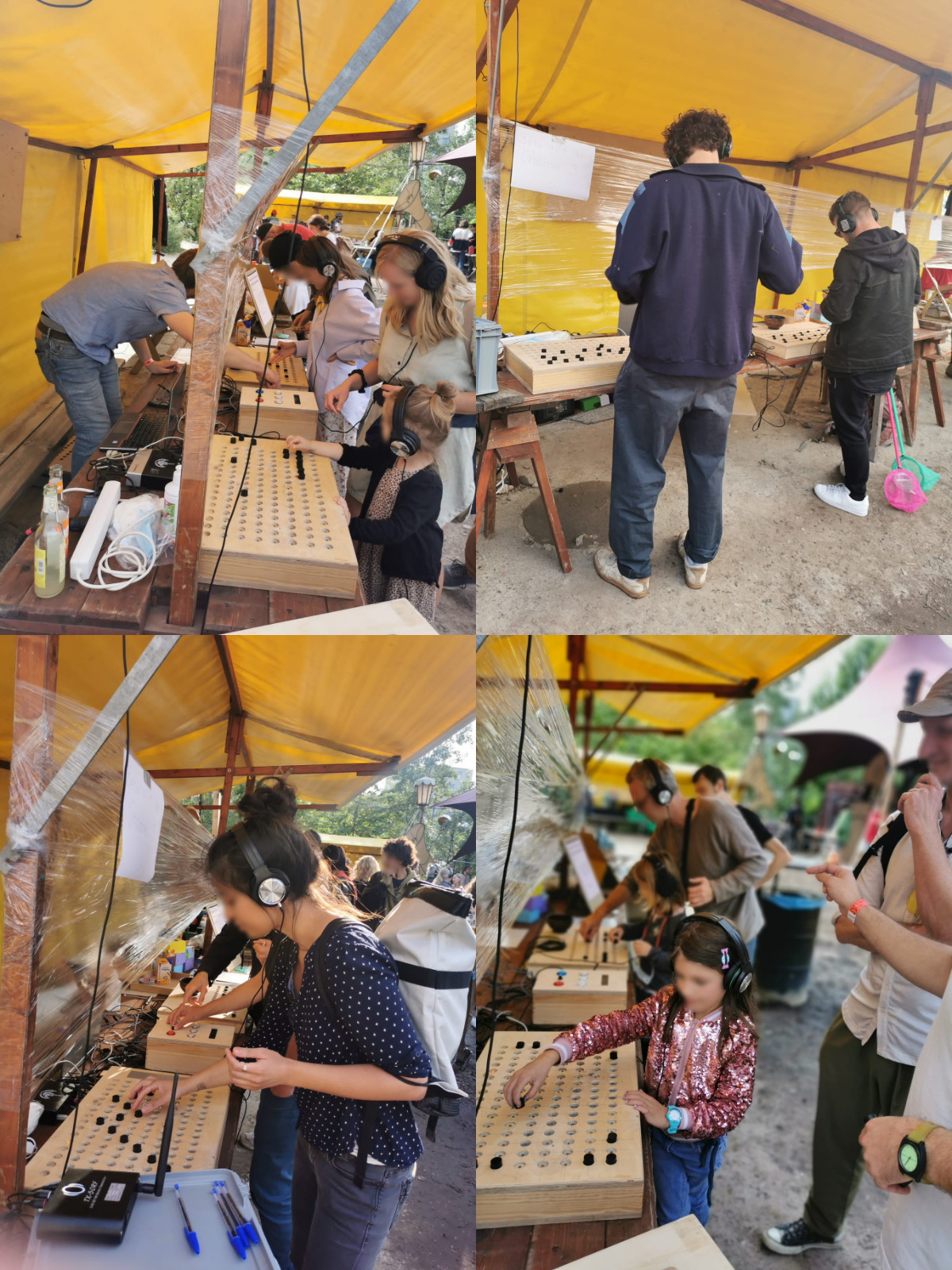}
  \caption{People interacting with LoopBoxes}
  \label{fig:testing}
\end{figure}

\section{Pilot Study with Mixed Audience}

The first evaluation took place during a public music festival (\textit{Festival für selbstgebaute Musik}\footnote{
\url{https://www.selbstgebautemusik.de/}
}) in Berlin, Germany, in an informal ``market-like'' setting outside (see figure \ref{fig:testing} for the setup). This festival aims to be an inclusive place for adults and children to enjoy and explore all kinds of self-built musical instruments.

We invited visitors of the festival to try out the instruments and collected feedback based on observation and a short questionnaire. We were particularly interested in how users experience the functionality of the instrument, to see if the instrument is perceived as aesthetically appealing to a general public, and to receive feedback on our design. We also wanted to evaluate if the instrument is suited for music making without the need of prior training and if the instrument is suited to foster a collaborative music making process.

For the evaluation, we designed a short questionnaire (see Figure \ref{fig:feedb}), focused on the main design considerations of the instrument. The rating scale used to evaluate the nine aspects had two options: ``not at all'' to ``very much'' for item 1 or to ``completely'' for items 2 through 9. Additionally, we asked for an assessment of the suitable age span of target users and the respondents' age, gender, profession, and musical training as well as qualitative feedback. The questionnaire was available in German and English. In addition, we observed the people using the instruments and talked to many of them in person.

\subsection{Observations}
The participants who used the instrument ranged from very young children (approximately 3 years) to adults over 60. Most people used the instrument for 10 to 30 minutes. We observed families forming a `musical ensemble' as well as strangers engaging in a collaborative music process.

In some cases, the timeline metaphor seemed hard to understand for participants. Generally, this was the case for very young children, but also for some older children as well as adults, especially in the following scenario. When two modules were already being used and a new musician entered the musical ensemble, it seemed to be hard to grasp one's own part in the music. In some cases, we had to explain the functionality of the different modules to the users, but in most cases, we observed them explaining to each other how they perceived the functioning of the instruments.

We also observed the users communicating about aesthetics. For example, one user said that ``this one does not fit'' referring to a single block on the step sequencer or another user said ``no, this sounds muddy'' referring to the density of the musical outcome.
Furthermore, we observed some users of module three disturbing/interrupting the others by changing the background loop or using the stutter-effect extensively. 
One very young child tried to stack the blocks onto each other while using the melodic step sequencer and seemed to expect feedback on this interaction.

\subsection{Questionnaire -- Quantitative Evaluation}

We received completed questionnaires from 39 participants (19 male, 17 female, 1 diverse, and 2 unspecified gender indications). The age ranged from 3 to 61 years ($\bar{x} = 34.72$, $\sigma = 11.75$). Three children ($age < 10$) participated in the evaluation on their own (with permission of their parents), in other cases they were assisted by their parents. The professions varied broadly from (SEN) (music) teachers, (music) therapists, social workers, to designers and engineers, a baker, and others. Around half of our participants indicated no musical training ($n =17$) and the others half indicated private musical experience ($n = 15$) and/or professional musical experience ($n = 7$).

\begin{figure*}[htbp]
  \centering
  \includegraphics[width=0.9\textwidth]{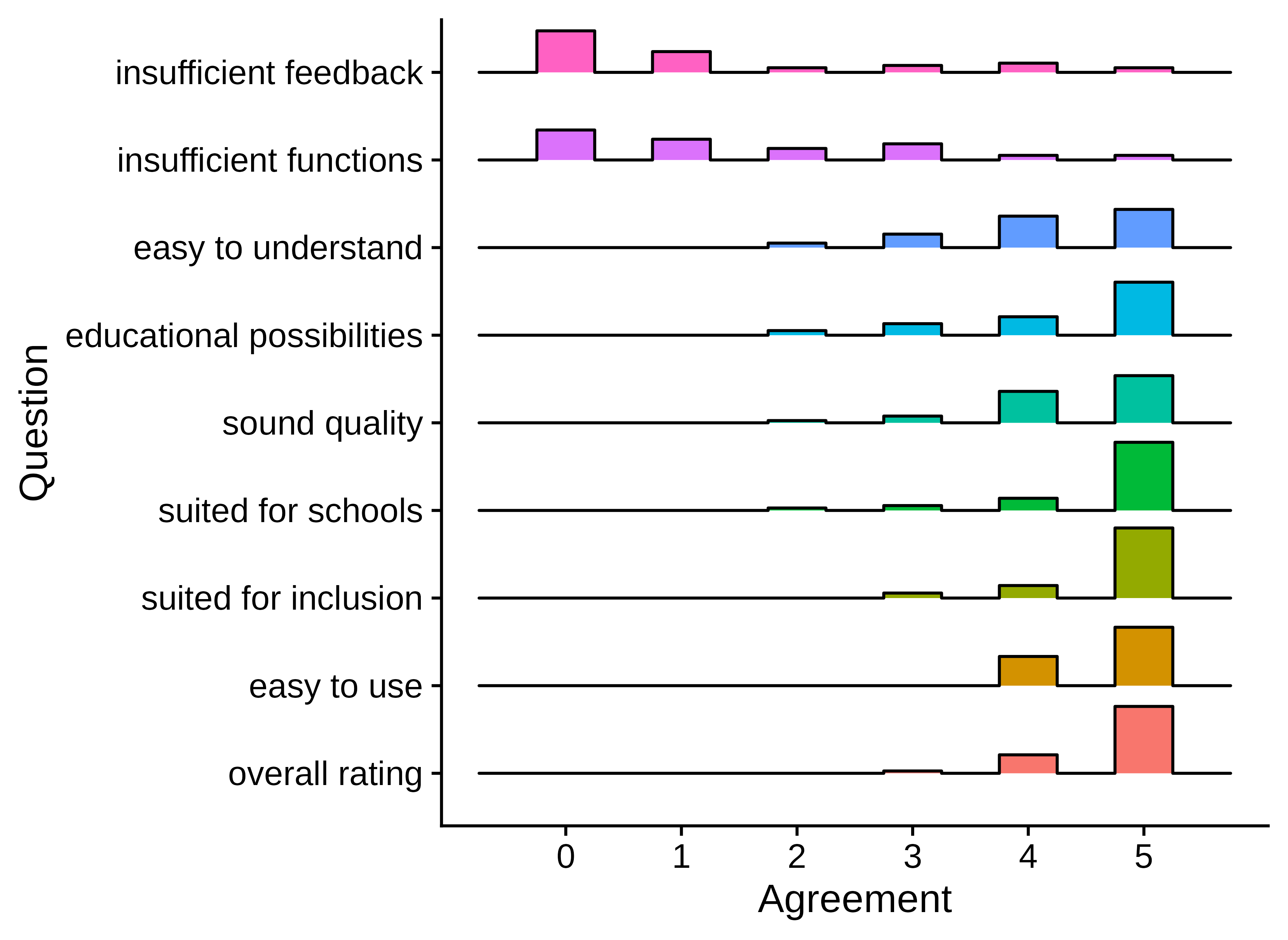}
  \caption{Density plot of the agreement to individual items by respondents on a scale from 0 - not at all - to 5 - completely}
  \label{fig:ratings}
\end{figure*}

Figure \ref{fig:ratings} shows the rating distribution. For the reversed items 4 and 7, the mean rating is smaller than 2. For all other items the mean rating is greater than 4.

The minimum age recommendations for the target user group according to our participants varied between 2 and 12 years ($\bar{x} = 5.61$, $\tilde{x} = 5$, $\sigma = 2.6$). The maximum age recommendations varied from 10 to 109 years (excluding one outlier at 180); ($\bar{x} = 74.20$, $\tilde{x} = 99$, $\sigma = 38.40$). Most of the participants indicated that the instrument is not only suited for children, but also for adults ($n = 5$ participants recommended a maximum age below 18). 

\subsection{Questionnaire -- Qualitative Evaluation}

Table \ref{tab:qual} provides an overview of the indicated positive aspects, negative aspects, and requests formulated in the questionnaire.

\subsubsection{Positive aspects}
\begin{table}
\scriptsize
\centering
  \caption{Requests, positive and negative aspects as indicated by the respondents}
  \label{tab:qual}
  \begin{tabular}{lll}
   
    \toprule
    positive & negative & requests\\
    \midrule
    tangibility & experimental effects & vocal inclusion\\
    direct feedback & harmonic domination & infographic\\
    no screen & lack of volume control & stackable blocks\\
    ease of use & & DIY kit\\
    reduction&&\\
    design&&\\
    discovery learning&&\\
    collaboration &&\\
    communication&&\\
    sound quality&&\\
    robustness&&\\

  \bottomrule
\end{tabular}
\end{table}


The overall feedback was very positive. Participants indicated that they enjoyed the haptic qualities of the interaction. A SEN music teacher wrote that the tangibility supports the comprehension of the different components of a beat. In this regard, one participant also mentioned the absence of a screen as a positive aspect.
The direct feedback seems to be strongly related to the evaluation of the instrument as being easy-to-use. Furthermore, the participants mentioned minimal functions, simplicity, a game-like-approach and the aspect that ``no matter what you do, it somehow sounds good'' in this context. One participant wrote that the instrument enables `all' people to experience music-making.
Regarding the design, the combination of visual feedback with light and wood as a building material was mentioned and seen as positive, and so was the out-of-the-box functionality.
Some participants wrote about the possibilities of fostering musical learning activities with the instrument, in the sense that different rhythms can be tested easily in conjunction with a direct experience. Additionally, the visualization of abstract concepts like harmony and rhythm was mentioned. The SEN music teacher claimed that the possibility to remove the parts that are not supposed to sound make the instrument concrete and suitable for people with learning difficulties.
Another aspect the participants enjoyed was working with others, including communication about and through music, opportunities for interaction, teamwork and generally getting along with different people.

\newpage
\subsubsection{Negative aspects}

On the negative side, one participant found module three ``too experimental'' and ``messy'', proposing more simple effects like ``delay, reverb, echo'' as opposed to the stutter effect provided by the instrument. This participant also proposed the need for more detailed volume controls for the drums and the melody.
Another aspect that was criticized is the harmonic domination of the instrument.

\subsubsection{Requests}

Two participants proposed the extension of including a way to record `your own voice' to the instrument. Furthermore, a do-it-yourself (DIY) construction kit was desired as well as the possibility to choose different sounds and to have some infographics explaining the instrument. The youngest participant (age = 3) proposed the integration of stackable blocks as an interaction possibility.

\subsection{Intermediate Reflection}

The feedback from the pilot study was predominantly positive. All items of the questionnaire had a strong tendency to their positive pole, although the inverted items (4 and 7) had more variance including some negative ratings. The qualitative feedback corresponding directly to the design requirements formulated by \cite{forsterhci, frontiers23} suggests that, overall, LoopBoxes have successfully met the needs from the respondents point of view.

\begin{figure}[t]
  \centering
  \includegraphics[width=.8\linewidth]{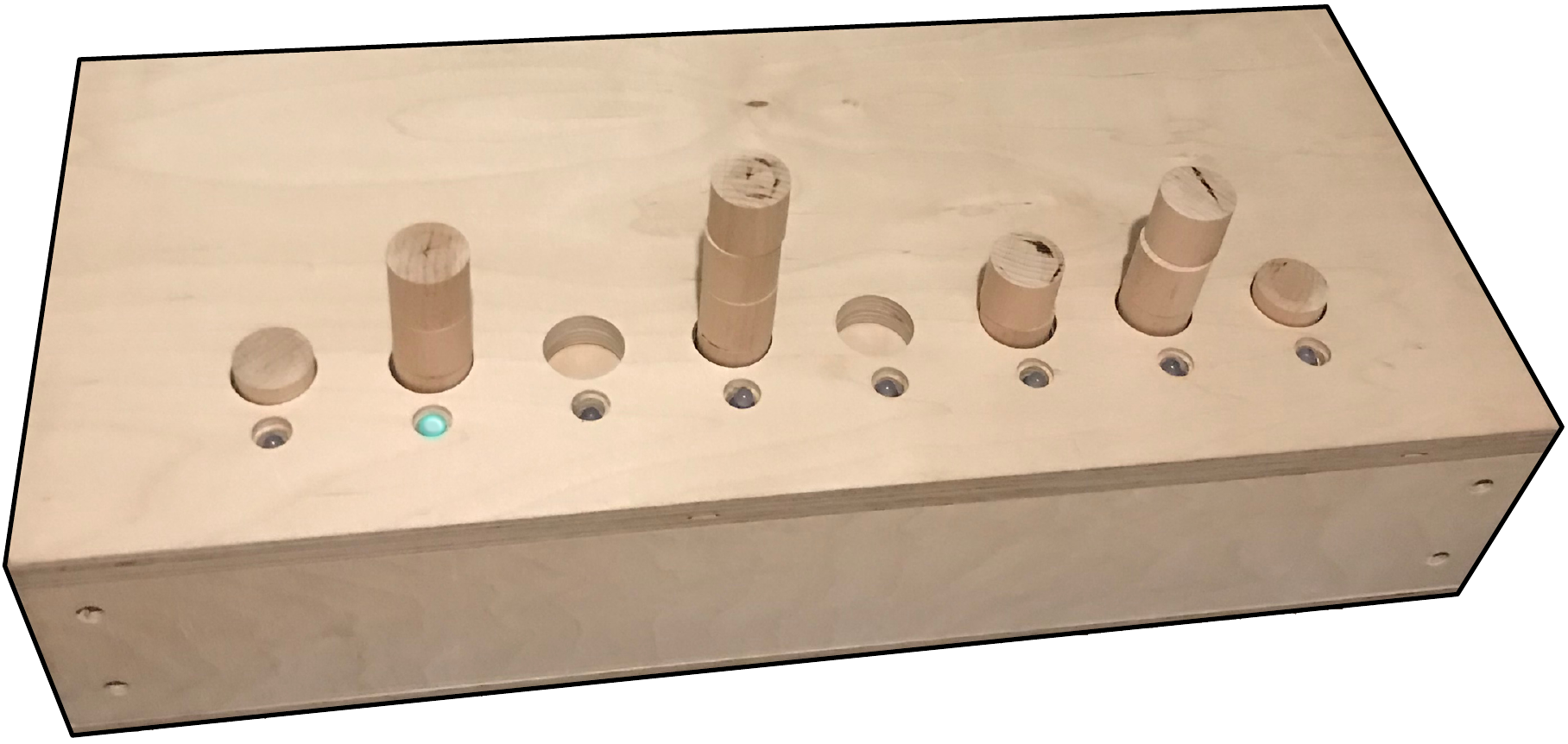}
  \caption{Improved iteration of the melodic sequencer.}
  \label{fig:melody}
\end{figure}

The least positive rating on the other scales is item 2 - `easy to understand', which can be explained by the fact that the timeline metaphor requires abstract cognitive skills and some level of experience interacting with a step sequencer. 
As a consequence, we reduced the melodic sequencer to one single row that uses a block-stacking-approach to determine note pitch (see figure \ref{fig:melody}). This idea is also based on our observation of one child intuitively stacking blocks on top of each other. 

\section{Case Study in SEN School Setting}

The second field test was a case study of the instrument with children and music teachers in a SEN school. The connection with the school was established through an ongoing participatory design project, conducted by the first author. The evaluation was based on observations during a class where the children made music together, and a semi-structured interview (appr. 60 minutes) with one music teacher.
 

The project started in September 2022 as a weekly music workshop, for which the children could register voluntarily. We deliberately refrained from collecting precise diagnoses on the children's disabilities in order to focus on their abilities. Overall, this was a group with highly heterogeneous abilities in the physical, cognitive and communicative domain. Nevertheless, for the discussion of results it seems necessary to differentiate the observations with regard to the severity of the disability, since students from the so-called group of people with complex disabilities (profound disabilities in all of the domains mentioned above) in particular experience access barriers. A total of eight children participated in the project, four of whom are accompanied by an individual assistant and communicate non-verbally.

Each session started with a common welcome song, after which the children were invited to try out an instrument. After making music together, we listened to music of different styles, which were evaluated by the children according to their individual preferences (good, neutral, bad). The indicated preferences were used to select appropriate musical styles for the instrument. We ended each session with singing a song. 

Since most of the students communicated non-verbally or only to a limited extent verbally, observation and image-based forms of communication were used for evaluation. The observations were discussed with the individual assistants as proxies during the sessions as well as with one music teacher at the end of each session. 
The second teacher was involved in the evaluation using a qualitative semi-structured interview.

It became apparent early on that much more time than originally planned would be needed to give the children enough space to explore the instruments. Therefore, only the drum sequencer and the loop controller could be tested so far. The effect slider has also not yet been used. However, the melodic sequencer has been presented to the teachers and reflected on as part of the interview.

\subsection{Results and First Conclusions}

Since we worked with a highly heterogeneous group, some children were able to operate all the instruments' modules without difficulty, while for others especially the drum sequencers' abstract form of interaction as well as its fine motor skill requirements did not seem to be appropriate. Furthermore, some students exhibited exploratory behaviors with the instruments on their own, while others seemed to expect more direct instruction by the teachers.

\begin{figure}[t]
  \centering
  \includegraphics[width=.8\linewidth]{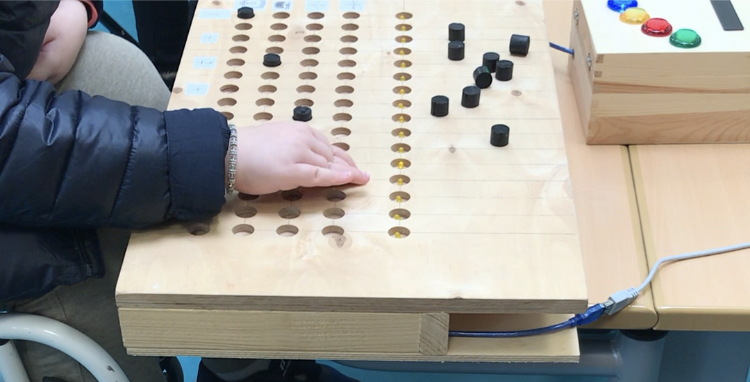}
  \caption{A student exploring the step sequencer module}
  \label{fig:loop_test}
\end{figure}

\begin{figure}[t]
  \centering
  \includegraphics[width=.8\linewidth]{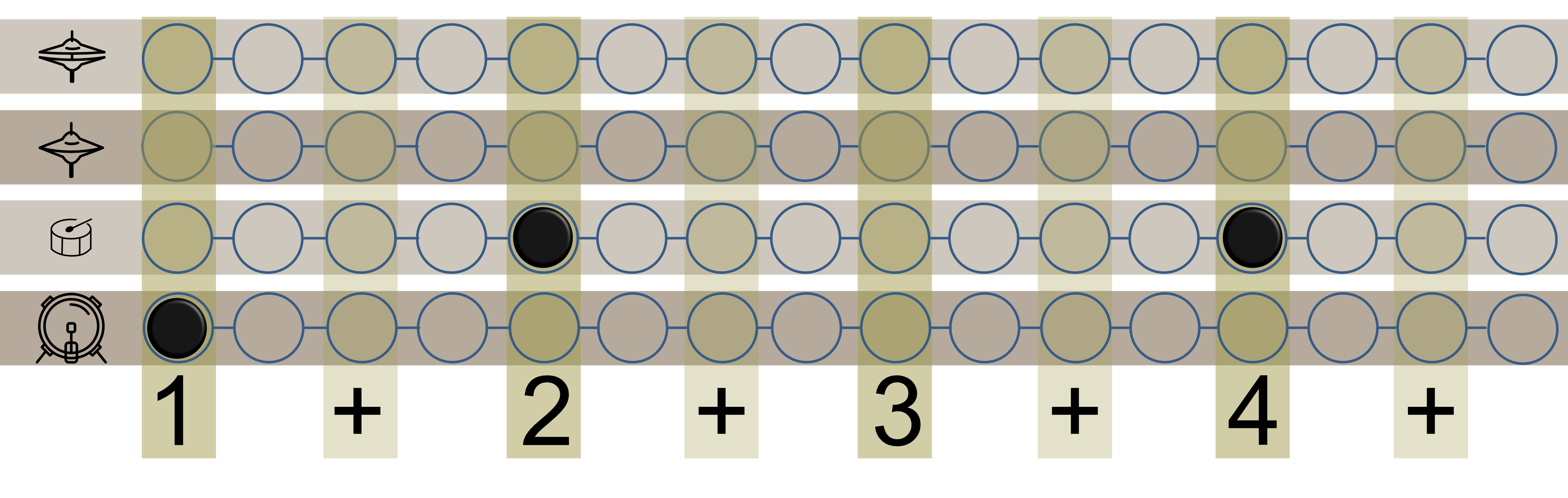}
  \caption{Example of graphic notation for drum sequencer}
  \label{fig:rhythm}
\end{figure}

\subsubsection{Independent exploration and graphic notation}
Initially, we gave the students the opportunity to independently explore the functions of the instrument's separate modules. 
For example, one student first placed one block in each row of the drum sequencer spread out horizontally, exploring the different sounds (see Figure \ref{fig:loop_test}). Another student placed a snare sound on the first beat, so the rhythm was perceived to be shifted by one beat. This student explored different positions for the snare drum and bass drum by placing one block and then waiting to listen to the feedback. For those students, the timeline metaphor seemed easy to understand and the interaction with the instrument itself seemed to be accessible. 
After the initial exploration, we offered graphic notation of rhythms as support (see Figure \ref{fig:rhythm}), which could be used as a starting point for exploration. 
In this regard, we observed that one student counted off the rows incrementally as she tried to copy the rhythms. Therefore, we added visual support to the surface of the drum sequencer in the form of vertical lines and numbers.

\subsubsection{Challenges and opportunities in interaction}
Especially for students with complex disabilities, the interaction of the instrument seemed to be less suitable. We observed children experimenting with different (not our intended) forms of interaction. 
For example, one student tapped on the wood while exploring the instruments or hit several lighted buttons with his hand on the loop module in a rather percussive way. He seemed to particularly enjoy the direct haptic interaction, but not care for the sonic feedback from the instrument. This student always carried a rattle in the shape of a ball in his hand.
This impression was also confirmed by the interviewed music teacher, who described the drum sequencers as difficult for students with complex 
disabilities.
At the same time, for all the other students, he describes the experience as a ``totally positive effect'' in which students ``form a work of art and create music in the process.''
In this context, he cites a ``higher creative part'' as an advantage of the drum sequencer compared to the loop sequencer, distinguishing between a conscious and ``random'' creative performance. But he also describes the unintentional creation of a rhythm (i.e., to simply put wooden blocks without musical intention) as an advantage, since it enables students who do not (yet) have a feeling for rhythm to engage in basic creative processes. He emphasizes that it is difficult to understand whether or not some students perceive themselves as sound producers in this context, but that he assumes they understand in ``some way'' their influence on the sound. He sees another advantage for children who generally have difficulties keeping a rhythm due to motion control limitations. He described the fine motor requirements as a main disadvantage, although he contrasts this with the advantage of the melodic sequencer, as the interaction is more accessible due to the larger size.

The loop module could be operated by all students. However, as already described in the last example, not all students seemed to perform the interaction with the intention to interact with our intended musical affordances. 
One student rejected the interaction with the loop module by pushing the instrument away from herself, but then wanted to put wooden blocks in the drum sequencer. It remained unclear whether she perceived herself as a sound producer.
Regarding the touch-interaction, the interviewed teacher highlighted the aspect that, regardless of motor limitations, the instrument affords the creation of nice sounds. He described that as ``a great experience''. In addition, he describes the ``big illuminated buttons'' as very attractive to students with complex disabilities.

\subsubsection{Collaborative music making}
Making music collaboratively using both modules combined seemed challenging for most students without appropriate support.
For example, we observed one student arranging different combinations of loops while a beat was laid on the drum sequencer. Then she stopped interacting for a longer period of time and it seemed like she did not know what to do next. As the beat was slowly reduced on the drum sequencer by taking away blocks, she began to also turn off the loop layers one by one, creating an effect resembling of a fade-out. 
This example shows that some students need more didactic support to engage in the exploration of collaborative music making processes. Besides, the instrument should be extended by further modules that offer direct haptic interaction as well as direct musical feedback. 

The interviewed teacher described the modular design of the instrument as an overall advantage. He cites the example from ``regular'' music classes, where students with complex disabilities in particular are mostly excluded from active music making or participation is limited to simple instruments like acoustic drums. According to him, a key advantage of the modular design of our instrument is that the modules complement each other in ``a certain way'', offering each student participation in the overall musical result which is much greater compared to the use of traditional instruments. The instrument is therefore particularly suitable for enabling very heterogeneous groups to actively participate in making music together. 

In this context, he also describes the advantage that by involving all students in making music, a change in perception of individual students by other students would be achieved. Hereby he addresses  an improvement of social participation. 

\subsubsection{Benefits of individual exploration}
Another observation we made in relation to the use of the instrument in group settings is the example of a student who at times showed basically no interest to participate in trying out the instrument. For this student, it proved to be a viable solution to first explore the instrument in an individual setting. In a conversation with the teacher, it was confirmed that the barrier did not lie in the instrument itself, but rather in a general reticence of the student in social situations.
For students with complex disabilities, the group setting does not seem to be ideal either, as it is more difficult to perceive their reactions and they often seemed easily distracted by other students. 

\subsubsection{Requirements for long-term instrument use}
In order to use the musical instrument in the longer term, the interviewed teacher addresses the need to reduce ``technical dependency'' and preparation time. Following him, it is important that the instruments are ready to use right away without any preparation required, like a keyboard that can be turned on with a switch. 
In addition, he mentioned the possibility of involving students in the construction of such instruments within longer-term school projects.

 \section{Discussion and Conclusion}
 
\subsection{Overall Evaluation of the Instrument}

Both the pilot study and the case study showed that the instrument met our design criteria for many users, but also highlighted aspects that should be improved or expanded in future iterations. Teachers judged the instrument to be a good way to engage all students in music lessons, including those who are normally excluded from active music making during music class, allowing them to receive a significant role in a collaborative musical work. This suggests that the instrument can increase access to aesthetic experiences as well as social participation in the music classroom. 

\subsubsection{Expanding the instrument's modules}
It was equally apparent that not all interaction possibilities are accessible or appealing to all students. However, the modular structure affording various interaction possibilities was described as a decisive advantage. Therefore, our goal is to expand the existing modules with additional ones. The design approach of having many simple instruments, from which the appropriate ones can then be chosen according to individual needs, is preferred by most SEN music teachers \cite{frontiers23}.
For students with complex disabilities, button interaction was evaluated as being the most accessible, and according to the teacher's assessment, also very appealing. But it remained unclear whether the students perceive musical value for themselves. 
Therefore, one of the things we plan to do is to design a ball as an instrument that affords music making through movement, based on our observation of a student who regularly interacts with a similar object.  

\subsubsection{Alternative musical styles and voice inclusion}
It turned out that for use in the SEN area, a lot of time needs to be allocated for instrument exploration. Therefore, some functions of the instrument and suggestions from the pilot study could not yet be considered -- e.g., proposals like the reduction of harmonic domination.
However, the musical style of the loops offered was rated well by the students. Nevertheless, in the future we are planning to offer also alternative styles and let the children decide what they like most. 
Also, voice inclusion was prepared by adding a recording function to the pure data patch of the sequencer and will be tested in the future.

\begin{figure*}[htbp]
  \centering
  \includegraphics[width=.8\linewidth]{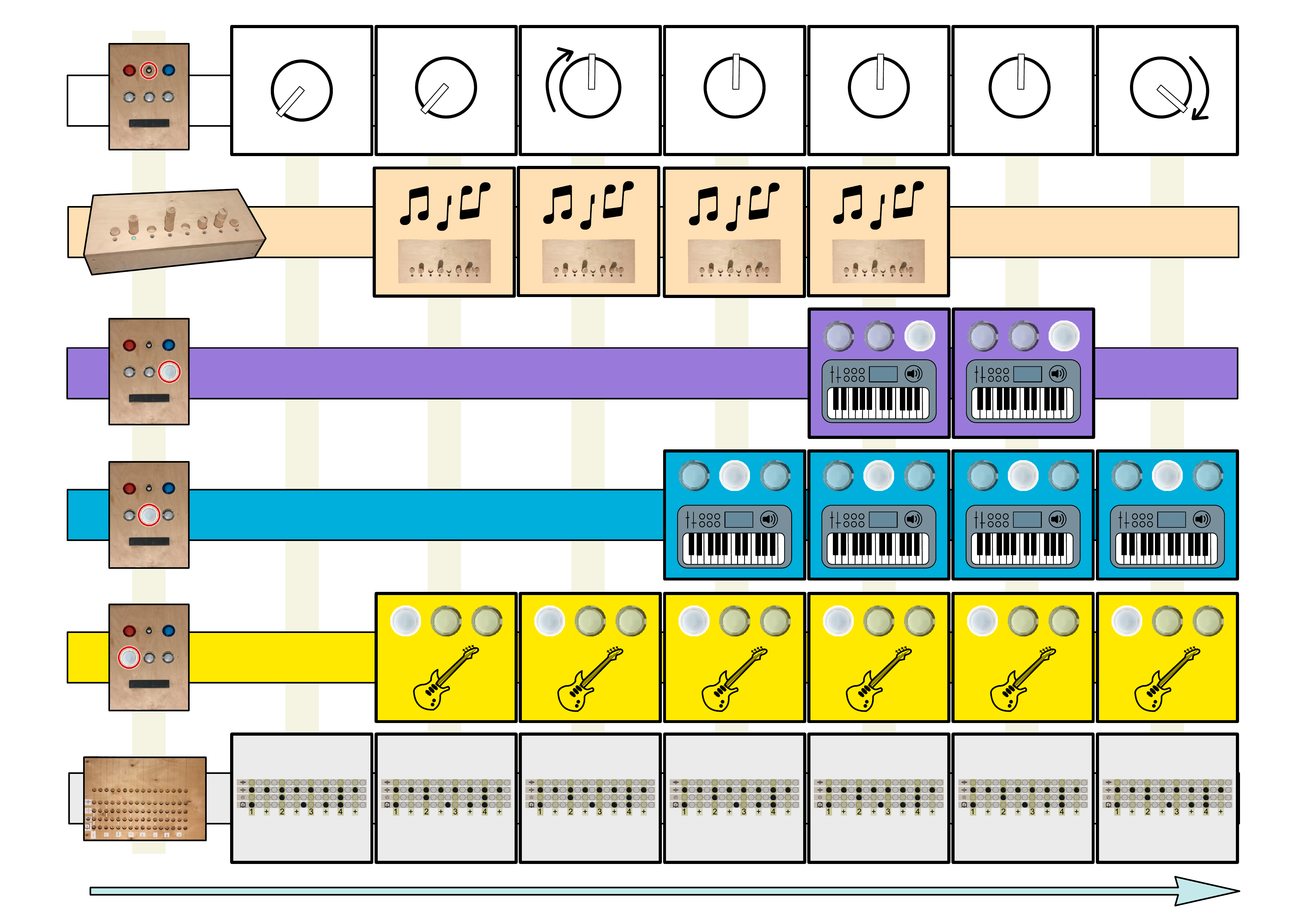}
  \caption{Graphical notation as an example of a possible song structure}
  \label{fig:partitur}
\end{figure*}

\subsubsection{Group vs. individual exploration}
It also became apparent that exploration in a group setting is not the best choice for all children.
There were students who disliked being the center of attention and therefore preferred to test the instrument alone first. 
Other children were quickly distracted by different stimuli. Therefore, an ideal approach would be to offer exploration in individual as well as group settings in each session, which unfortunately is only feasible to a limited extent due to the requirements of the school. Nevertheless, if possible, we plan to offer additional individual sessions, especially for the group of students with complex disabilities.

\subsubsection{Scaffolding for collaborative music making}
Specifically, to facilitate collaborative music making practices, it seems necessary to provide didactic material as scaffolding for most of the students. 
For example, the interaction with the loop module seemed accessible to most children, but the combination of different musical layers was an unfamiliar practice. Therefore, didactic materials (see Figure \ref{fig:partitur}) were developed for future sessions in the form of graphic notation scores that may be used to visualize or facilitate the composition of musical sequences.

\subsubsection{Aesthetic communication}
So far, we have not been able to observe any aesthetic communication \cite{wallbaum} among the children, which is probably due to the children's predominantly non-verbal communication abilities and to the discussed need for more direct instruction. 
The pilot study showed that aesthetic communication might be stimulated as soon as the users are sufficiently familiar with the instruments' functions. In the SEN context, this probably requires a lot of preparation time and more support.

\subsubsection{Flexible and easy-to-use technology}
Most important for longer-term use in schools, as indicated by the teachers and in \cite{davis}, is enabling the instrument to function independently and being ready for use without extended preparation time. The idea is to make every single module run on a raspberry pi, sync via wifi and to be controllable via a web-interface, so that it is easy to use and remains flexible at the same time. 

In the pilot study, the desire for a do-it-yourself-kit building set was expressed. This aspect was also addressed by the teacher in the interview. Here, one goal of our design from wood in conjunction with open source technology was to enable such projects.

\subsection{Summary and Future Prospects}

Overall, the results of the pilot study and case study suggest that the instrument was successful in meeting many of the design criteria and was able to engage all students in music lessons, including those who are typically excluded from active music making. The modular structure of the instrument was also seen as a major advantage and allowed for flexibility in choosing the appropriate modules for individual needs. However, it was also apparent that not all interaction possibilities were accessible or appealing to all students, particularly those with complex disabilities. Therefore, future iterations will focus on expanding the modules to better serve this population. 
Since the instrument was enjoyed by children as well as adults ranging from formally trained musicians to hobby musicians and people with no musical experience during the pilot study, we hope that it could also be aesthetically appealing to children in inclusive settings, facilitating collaborative music-making between children with and without disability experience in the long term. 

\newpage

\section{Acknowledgments}
We thank \textit{Aktion Mensch e.\,V.} for supporting our project with a financial donation and \textit{ueberschall} for providing different musical loops. 
We also thank the participants of the pilot study as well as the children and music teachers at our partner school. Furthermore, we thank Achim Reinhardt for his support in the development of the improved melodic sequencer.

\section{Ethical Standards}
All persons who participated in our evaluation did so voluntarily. 
The children's parents were fully informed about the project and their rights and gave written consent for their children to participate. 
In addition, children were free to participate or not participate in each session. Especially with children who communicate non-verbally, their individual assistants help interpret whether or not a child is willing to participate or not.
Both the teachers and the school administration were fully informed about the project and their rights and gave their written consent.

\bibliographystyle{abbrv}
\balance
	\bibliography{nime-references} 

\appendix
\section{Questionnaire}
\begin{figure*}[htbp]
  \centering
  \includegraphics[width=1.3\linewidth, angle=90]{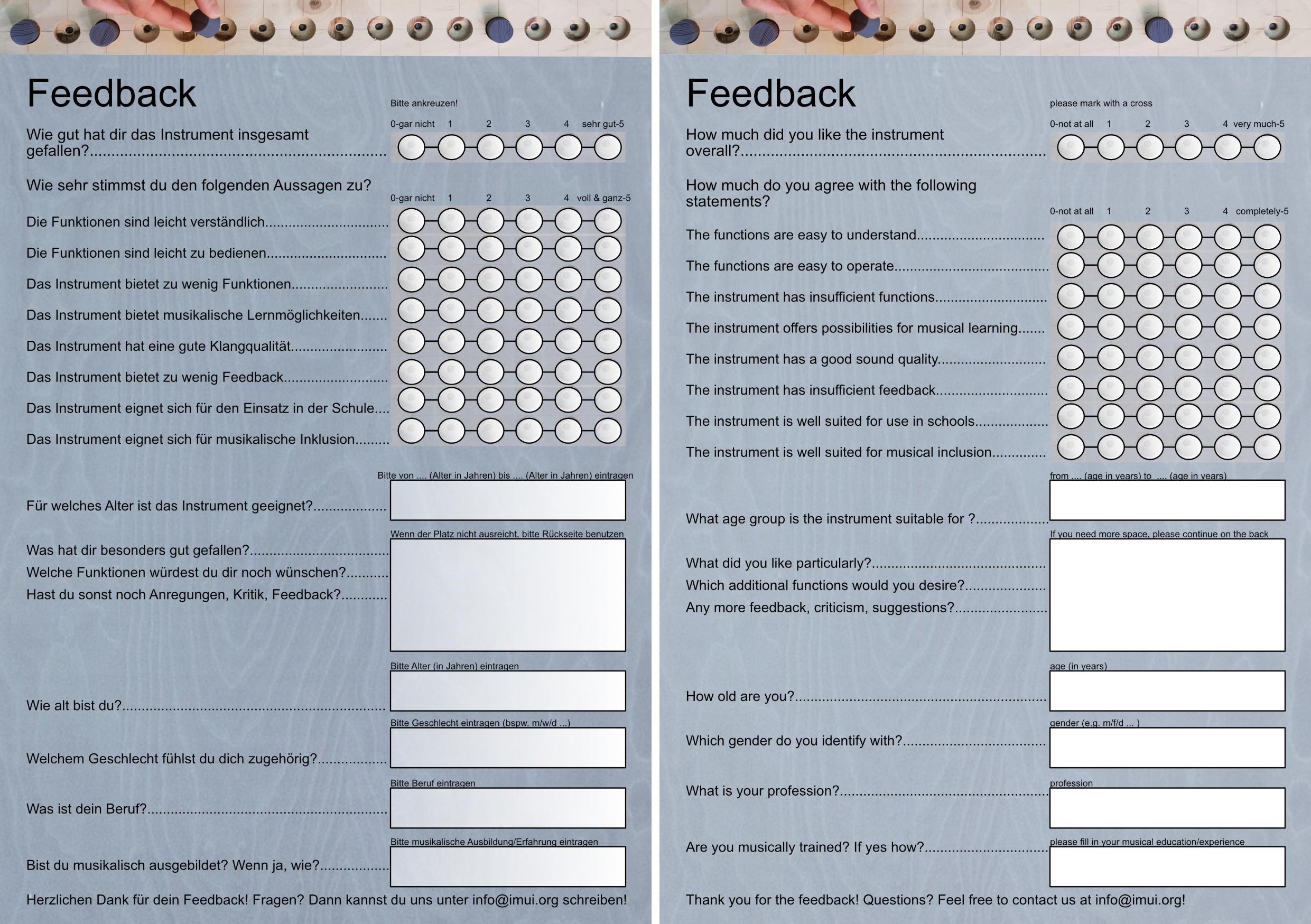}
  \caption{Questionnaire in German (left) and English (right)}
  \label{fig:feedb}
\end{figure*}
\end{document}